\documentclass[a4paper,12pt,aps,pra,superscriptaddress,reprint,floatfix,showpacs]{revtex4-1}

\usepackage[utf8]{inputenc}
\usepackage[english]{babel}
\usepackage[T1]{fontenc}
\usepackage{amsmath,amssymb}
\usepackage{graphicx}
\usepackage{braket}
\usepackage{amsfonts}
\usepackage{array}
\usepackage{lmodern}
\usepackage{esvect}
\usepackage{fancyhdr}
\usepackage{xcolor}
\usepackage{textcomp}
\usepackage[squaren,Gray]{SIunits}
\usepackage{natbib}
\usepackage{hyperref}
\usepackage[caption=false]{subfig}

\graphicspath{{Images/}}

\begin{document} 

\title{Energy deposition from focused terawatt laser pulses in air undergoing multifilamentation}

\author{Guillaume Point}
\email{guillaume.point@ensta-paristech.fr}
\affiliation{Laboratoire d'Optique Appliqu\'ee, ENSTA ParisTech, CNRS, Ecole Polytechnique, Universit\'e Paris-Saclay, 828 boulevard des Mar\'echaux, 91762 Palaiseau cedex, France}
\author{Emmanuelle Thouin}
\affiliation{Laboratoire d'Optique Appliqu\'ee, ENSTA ParisTech, CNRS, Ecole Polytechnique, Universit\'e Paris-Saclay, 828 boulevard des Mar\'echaux, 91762 Palaiseau cedex, France}
\author{Andr\'e Mysyrowicz}
\affiliation{Laboratoire d'Optique Appliqu\'ee, ENSTA ParisTech, CNRS, Ecole Polytechnique, Universit\'e Paris-Saclay, 828 boulevard des Mar\'echaux, 91762 Palaiseau cedex, France}
\author{Aur\'elien Houard}
\email{aurelien.houard@ensta-paristech.fr}
\affiliation{Laboratoire d'Optique Appliqu\'ee, ENSTA ParisTech, CNRS, Ecole Polytechnique, Universit\'e Paris-Saclay, 828 boulevard des Mar\'echaux, 91762 Palaiseau cedex, France}

\begin{abstract}
Laser filamentation is responsible for the deposition of a significant part of the laser pulse energy in the propagation medium. We found that using terawatt laser pulses and relatively tight focusing conditions in air, resulting in a bundle of co-propagating multifilaments, more than 60 \% of the pulses energy is transferred to the medium, eventually degrading into heat. This results in a strong hydrodynamic reaction of air with the generation of shock waves and associated underdense channels for each short-scale filament. In the focal zone, where filaments are close to each other, these discrete channels eventually merge to form a single cylindrical low-density tube over a $\sim \unit{1}{\micro\second}$ timescale. We measured the maximum lineic deposited energy to be more than \unit{1}{\joule\cdot\rp\metre}.
\end{abstract}


\maketitle

\section{Introduction}

Laser filamentation is a propagation regime for ultrashort laser pulses in transparent media reached when their peak power exceeds a critical power $P_{cr}$, which is about \unit{5}{\giga\watt} in air at a wavelength of \unit{800}{\nano\metre}. It results from a complex interplay between the optical Kerr effect, nonlinear energy absorption due to multiphoton and tunnel ionization and plasma defocusing \cite{Couairon2007}. While propagating in this regime, the laser pulse is able to maintain a high intensity over several Rayleigh lengths. Filamentation can deposit a significant part of the pulse energy in the propagation medium. Plasma generation itself, together with plasma absorption, form the main channel for energy deposition. Stimulated rotational Raman scattering on air molecules can also account for energy depletion of the laser pulse \cite{Nibbering1997,Seideman2001,Kartashov2006}. This energy, initially stored as potential and kinetic energy of plasma free electrons and as rotational energy of air molecules, is eventually converted into gas thermal energy after plasma recombination and rotational thermalization, occurring over $\sim \unit{1}{\nano\second}$ \cite{Tzortzakis2000} and $\sim \unit{100}{\pico\second}$ timescales \cite{Chen2007}, respectively. The resulting heating can range from $\sim \unit{100}{\kelvin}$ \cite{Cheng2013} to more than \unit{1000}{\kelvin} \cite{Point2015}, depending on experimental conditions. As air thermal conductivity is low, the system relaxes by launching an outward-propagating cylindrical pressure and density wave, expelling matter from the center of the channel and leaving a low-density tube. This underdense channel then slowly evolves by thermal diffusion, which can take times up to several tens of milliseconds \cite{Point2015}. Such laser-induced air hydrodynamics are particularly interesting because they pave the way for the development of remotely generated, long-lived virtual optical structures such as optical waveguides \cite{Jhajj2014,Lahav2014}. They could also be used in the prospect of enhanced aircraft aerodynamics \cite{Dufour2013}. As the efficiency of these structures is strongly dependent on the amplitude of laser-induced hydrodynamics effects, that is on local deposited energy, it is crucial to be able to increase the density of deposited energy at the required level.

In this Article, we first study the influence of laser parameters on energy deposition. We found that using the shortest and most energetic pulse with a moderately strong focusing yields the best results overall, with more than 60 \% of the laser pulse energy being converted into heat at the terawatt peak power level. In this case the laser pulse peak power is well above $P_{cr}$, and the beam breaks down to form many co-propagating filaments due to modulational instabilities \cite{Mlejnek1999,Fibich2005}. This effect results in the pulse energy being deposited in competing short-scale filaments instead of being concentrated in a single structure, leading to an overall degradation of local energy deposition. Investigating the air hydrodynamics response by means of transverse interferometry, we witnessed a spontaneous evolution of the system from discrete underdense channels to a cylindrically-symmetric form with a single, large channel and an associated unique shock wave, partially negating the deleterious effect of multifilamentation. In optimal conditions, this channel lasts for more than \unit{100}{\milli\second}. Using a sonographic technique \cite{Yu2003,Point2015a}, we estimated the peak lineic deposited energy to be \unit{1.3}{\joule\cdot\rp\metre}.

\section{Optimizing energy deposition}

We first determined optimal experimental conditions to get the maximum absolute energy deposition from filamentation. To this purpose, we used a direct measurement of the laser pulse energy right after the focusing lens and $\sim \unit{10}{\centi\metre}$ after the end of the laser-induced plasma by means of a Joule meter (model QE50LP-H-MP from Gentec-EO). The difference between input and output energy was considered as deposited energy, which is a good approximation since energy losses due to Rayleigh and Thomson scattering are negligible. Such measurements were repeatedly done while varying laser pulse energy, laser pulse duration and focusing conditions.

\subsection{Laser pulse energy}

\begin{figure}[!ht]
\begin{center}
\includegraphics[width=.45\textwidth]{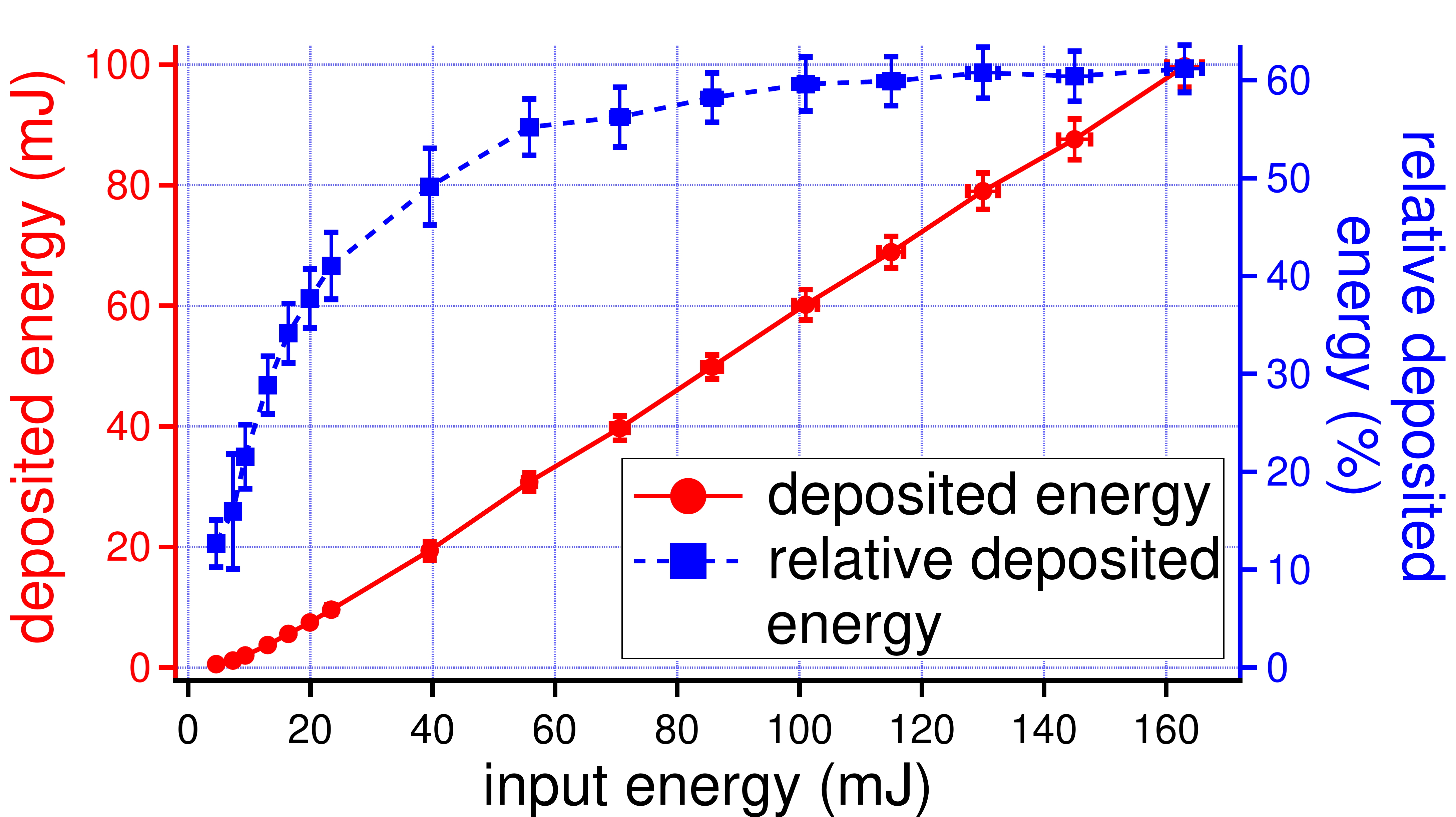}
\end{center}
\caption{(Color online) Evolution of the total (red circles) and relative (blue squares) deposited energy with input energy from the filamentation of a \unit{50}{\femto\second} laser pulse at \unit{800}{\nano\metre} focused at $f/30$. Error bars correspond to a confidence interval of one standard deviation evaluated over 500 shots.}
\label{figure_1}
\end{figure}

The first investigated laser parameter was laser pulse energy. For this study a moderately strong focusing at $f/30$ was used with a \unit{50}{\femto\second} laser pulse at \unit{800}{\nano\metre}. Evolution of deposited energy with input energy is displayed in figure \ref{figure_1}.

In this figure are plotted two curves: first, the total deposited energy (red circles), that is the difference between input and output energies and second, the relative deposited energy (blue squares), that is deposited energy normalized by input energy. The global trend is an increase of deposited energy with input energy. Deposited energy evolves quasi linearly at higher laser energies. As for relative deposited energy, it can roughly be divided in three parts: the first one, up to $E_{in} \approx \unit{20}{\milli\joule}$, is characterized by a very steep increase, reaching up to 35 \%. Past this point, relative deposited energy still increases, but much more slowly. Eventually, beyond $E_{in} \approx \unit{110}{\milli\joule}$, it saturates around 60 \%. We attribute the transition observed around \unit{20}{\milli\joule}, that is a peak power of about $80 P_{cr}$, to the threshold between the full beam self-focusing leading to an initial single filament generation and the local self-focusing inside the beam resulting in multifilament formation before the nonlinear focus, as described by Fibich and co-authors \cite{Fibich2005}. This means that as soon as the beam breaks up into short-scale filaments before the focal zone, where the majority of energy deposition occurs, increasing input energy leads to a redistribution of this energy over many competing filaments, only slightly increasing deposited energy for each of them. Conversely, as long as only one filament is concerned, any increase in energy will be channeled through this single structure and lead to a dramatically stronger energy deposition.

\subsection{Laser pulse duration}

Influence of the laser pulse duration on energy deposition was investigated by fixing input energy and focusing conditions and by detuning the laser compressor, imparting temporal chirp to the pulse.

\begin{figure}[!ht]
\begin{center}
\includegraphics[width=.45\textwidth]{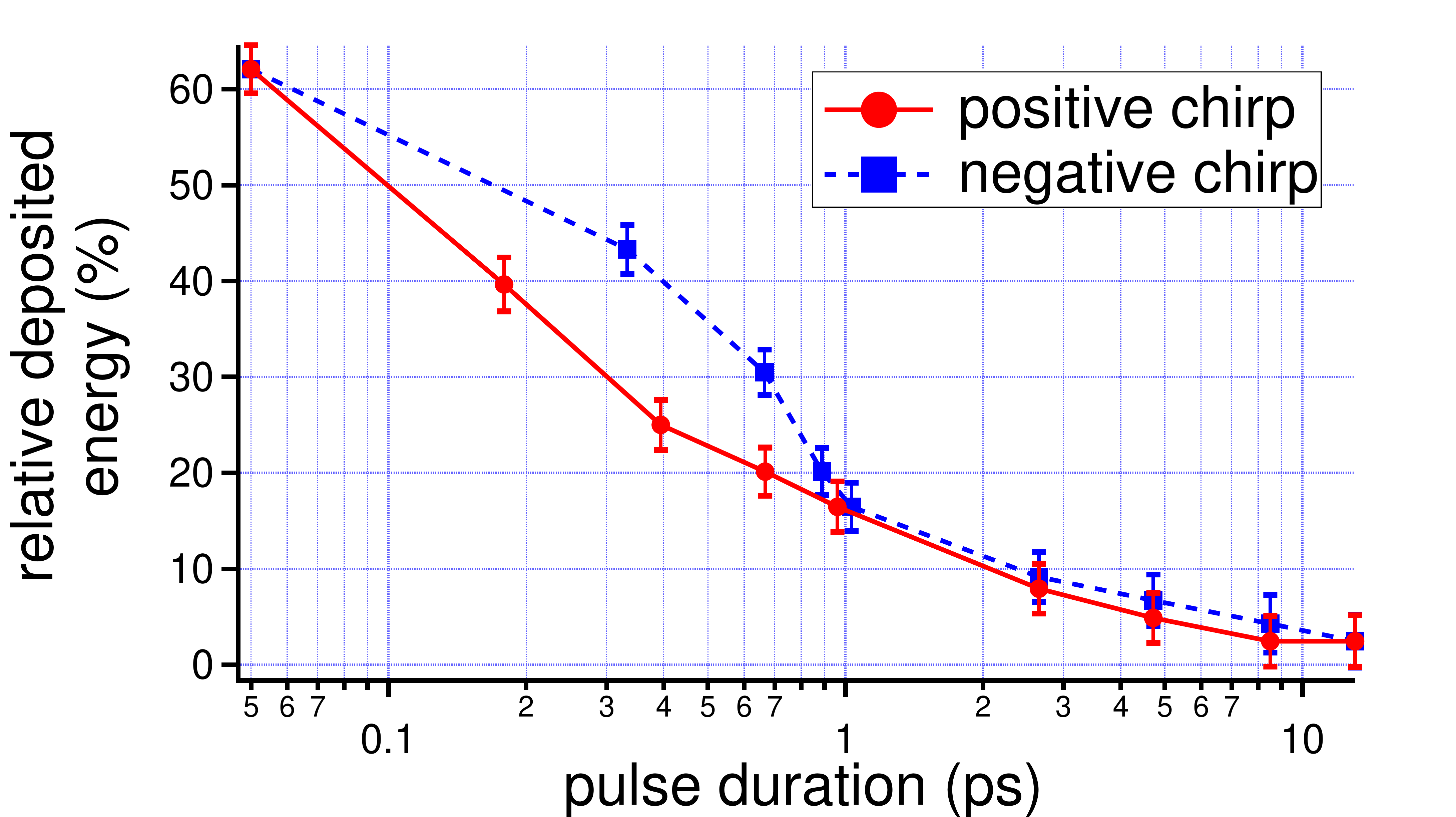}
\end{center}
\caption{(Color online) Evolution of the relative deposited energy with pulse duration in the case of positive (red circles) and negative chirp (blue squares). Filamentation was generated by a \unit{164}{\milli\joule} laser pulse at \unit{800}{\nano\metre} focused at $f/30$. Error bars correspond to a confidence interval of one standard deviation evaluated over 500 shots.}
\label{figure_2}
\end{figure}

Results displayed in figure \ref{figure_2} are unambiguous: pulse duration is a very sensitive parameter. As soon as it deviates from the minimum value of \unit{50}{\femto\second}, that is from a maximum peak power, deposited energy quickly falls, being reduced from more than 60 \% to less than 20 \% at \unit{1}{\pico\second}, and to only a few percents past \unit{10}{\pico\second}. It is worth mentioning that a negative chirp yields better results than a positive chirp. This can be explained by the plasma defocusing effect affecting the back of the pulse, which is stronger at longer wavelengths and will therefore tend to decrease the intensity of positively chirped pulses more than for negatively chirped pulses. With a long pulse duration (typically $> \unit{1}{\pico\second}$), plasma absorption due to inverse Bremsstrahlung, which is also more efficient at longer wavelengths, can also play a role \cite{Raizer1966}.

One could think that as some energy transfer channels, like avalanche ionization, are favored by longer pulses, it could positively affect energy deposition. Such behavior has indeed been observed during filamentation in fused silica \cite{Mermillod2008} and water \cite{Brelet2015}. Conversely, it seems that maximum intensity gives the best results in air, because it promotes a more efficient nonlinear ionization and Raman absorption, in the limit of pulse durations that can be achieved with our laser system.

\subsection{Focusing conditions}

\begin{figure}[!ht]
\begin{center}
\includegraphics[width=.45\textwidth]{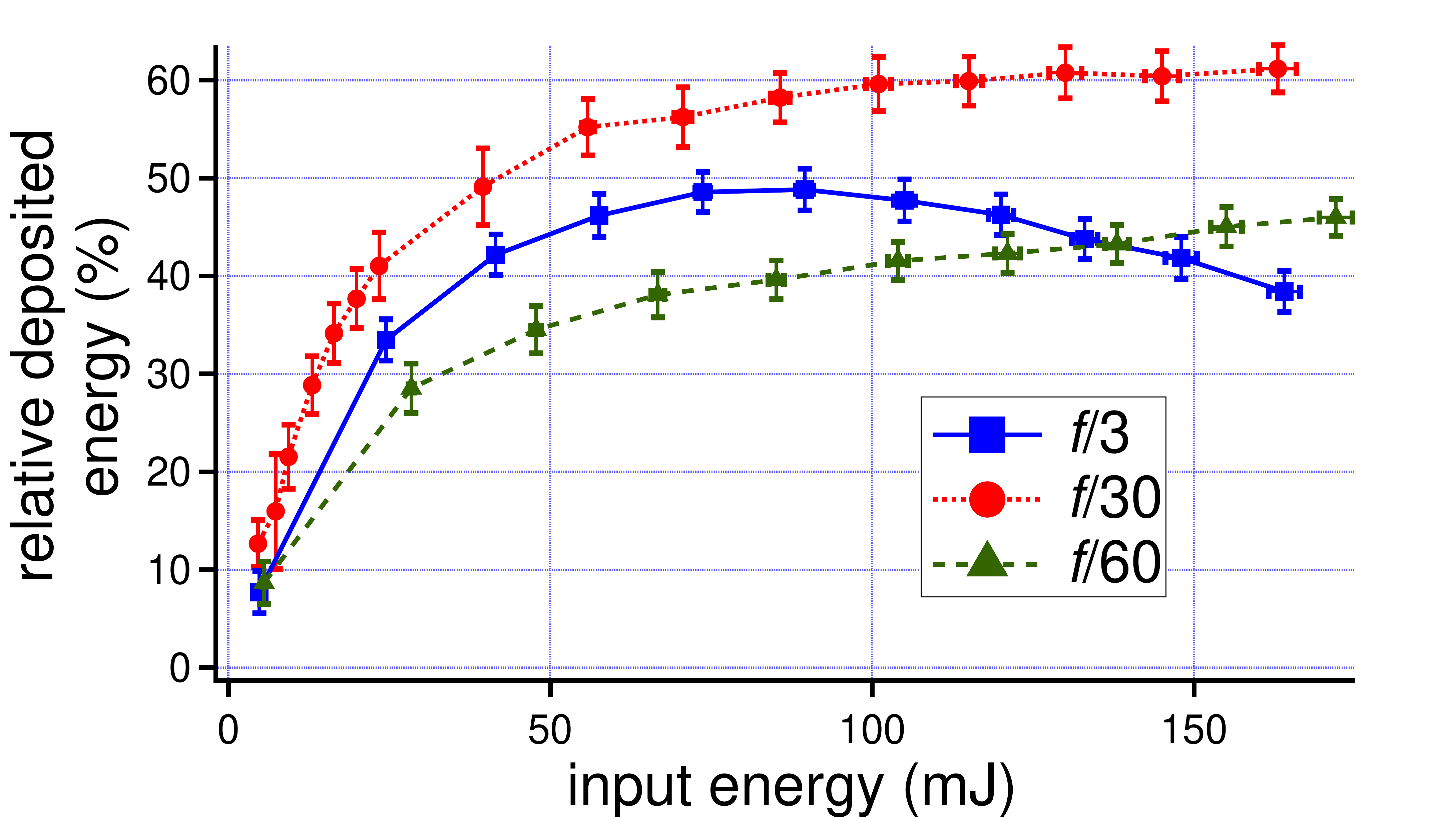}
\end{center}
\caption{(Color online) Evolution of the relative deposited energy with input energy using a $f/3$ (blue squares), a $f/30$ (red circles) or a $f/60$ focusing (green triangles). Filamentation was generated by a \unit{50}{\femto\second} laser pulse at \unit{800}{\nano\metre}. Error bars correspond to a confidence interval of one standard deviation evaluated over 500 shots.}
\label{figure_3}
\end{figure}

Influence of focusing conditions was also studied by comparing three cases: a moderate focusing at $f/60$, a stronger focusing at $f/30$ and an extreme focusing at $f/3$, corresponding to the quasi-breakdown regime described by Kiran \textit{et al.} \cite{Kiran2010}. Results are given in figure \ref{figure_3}. It shows that for a given input energy, energy deposition is maximum in the $f/30$ case. The three cases also exhibit a regime transition around \unit{20}{\milli\joule}, which is more obvious for weaker focusing, with a first phase where energy deposition quickly increases followed by a second phase where it rises slowly or stagnates.

Interesting features observed with the strong $f/3$ focusing are that deposited energy is significantly lower than in the $f/30$ case and that relative energy deposition reaches a maximum around \unit{85}{\milli\joule} at $\sim 50$ \% and then starts to slowly decrease, going below 40 \% at \unit{165}{\milli\joule}. This behavior might be explained by the fact that such a strong focusing promotes the generation of a very dense and short plasma with an accordingly strong defocusing effect, leading to a decrease of intensity and therefore to a less efficient energy deposition. Even though this decrease is modest, it results in energy deposition with $f/3$ focusing to be lower than energy deposition in the $f/60$ case at high input energy.

\section{Energy deposition in the high energy limit}

\subsection{Filamentation-generated hydrodynamics}

We then concentrated on the case that yielded the most important energy deposition: a \unit{165}{\milli\joule}, \unit{50}{\femto\second} (\unit{3.3}{\tera\watt}) laser pulse at \unit{800}{\nano\metre} focused at $f/30$. To characterize filamentation-induced hydrodynamics, we performed transverse interferometry on laser-generated air channels using the same instrument presented in references \cite{Point2014}, \cite{Point2015} and \cite{Point2015a}. 

\begin{figure}[!ht]
\begin{center}
\includegraphics[width=.45\textwidth]{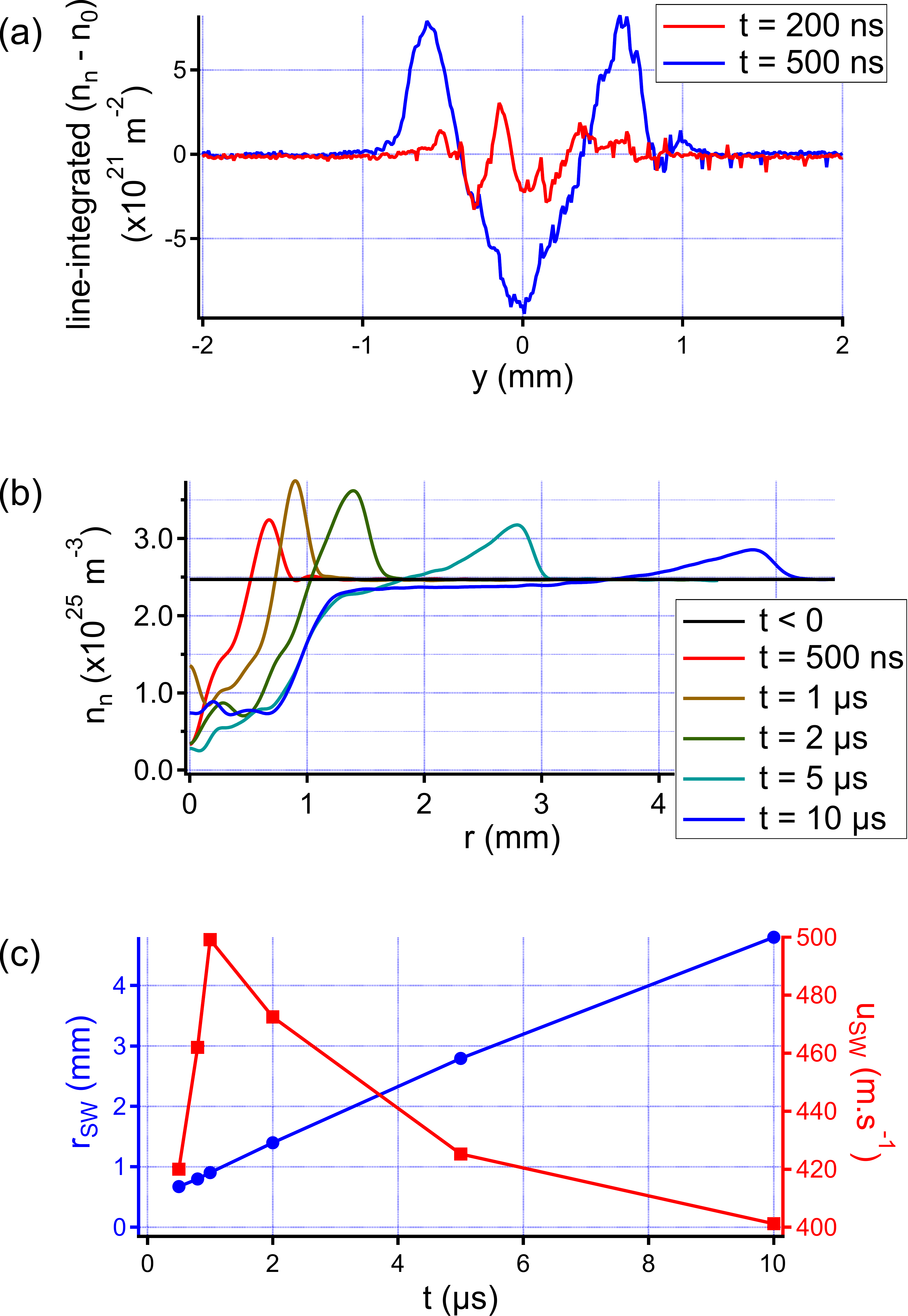}
\end{center}
\caption{(Color online) (a): line-integrated air density profiles at $z=\unit{1}{\metre}$ after the focusing lens (linear focus) for two different delays, displaying self-symmetrization. (b): radial air density profiles extracted after channel symmetrization at $z = \unit{1}{\metre}$. (c): evolution of the shock wave radius $r_{SW}$ (blue circles) and shock wave speed $u_{SW}$ (red squares). Filamentation was generated by a \unit{165}{\milli\joule}, \unit{50}{\femto\second} laser pulse at \unit{800}{\nano\metre} focused at $f/30$.}
\label{figure_4}
\end{figure}

As a large number of multifilaments is generated and, therefore, the laser beam exhibits a strong inhomogeneity, we did not expect the system to have a cylindrical symmetry, preventing us to use the Abel inversion to retrieve air density radial profiles. Nevertheless it was found that in the focal zone after a given delay, air density spontaneously evolves from a highly disorganized state, reminiscent of multifilamentation, to a cylindrically-symmetric state, as displayed in figure \ref{figure_4}-(a). In this last figure are plotted two line-integrated air density profiles taken at the linear focus. At delay \unit{200}{\nano\second}, several structures can be seen at the center of the profile, undoubtedly resulting from discrete short-scale filaments. The picture is very different at delay \unit{500}{\nano\second} because at this time, line-integrated density turned into a much more organized state with a good symmetry.

Since the system becomes cylindrically-symmetric after some time, Abel inversion can be used to compute air density profiles. Results are displayed in figure \ref{figure_4}-(b). These profiles are characterized by an outward-propagating cylindrical shock wave leaving a central underdense channel, much in the same way as single filamentation. In the present case however, the shock is broader and its amplitude is higher than in the case of high-energy monofilaments (see results given in reference \cite{Point2015} for comparison), while the underdense channel has approximately the same depth at about 30 \% of normal air density, but is significantly larger, reaching a full width at half maximum of \unit{2}{\milli\metre} after \unit{10}{\micro\second}. We still recorded the presence of the density hole after \unit{100}{\milli\second}, the maximum time we could reach due to experimental constraints. By this time, it enlarged so much that it affected the whole field of view of the camera, preventing us from extracting air density. Shock speed $u_{SW}$ was evaluated using shock radial position $r_{SW}$ and is plotted in figure \ref{figure_4}-(c). It exhibits a steep initial increase up to $\sim \unit{500}{\metre\cdot\rp\second}$ during the first microsecond, followed by a slower, gentle decrease at subsequent times, still propagating with a supersonic speed after \unit{10}{\micro\second}.

\begin{figure}[!ht]
\begin{center}
\includegraphics[width=.45\textwidth]{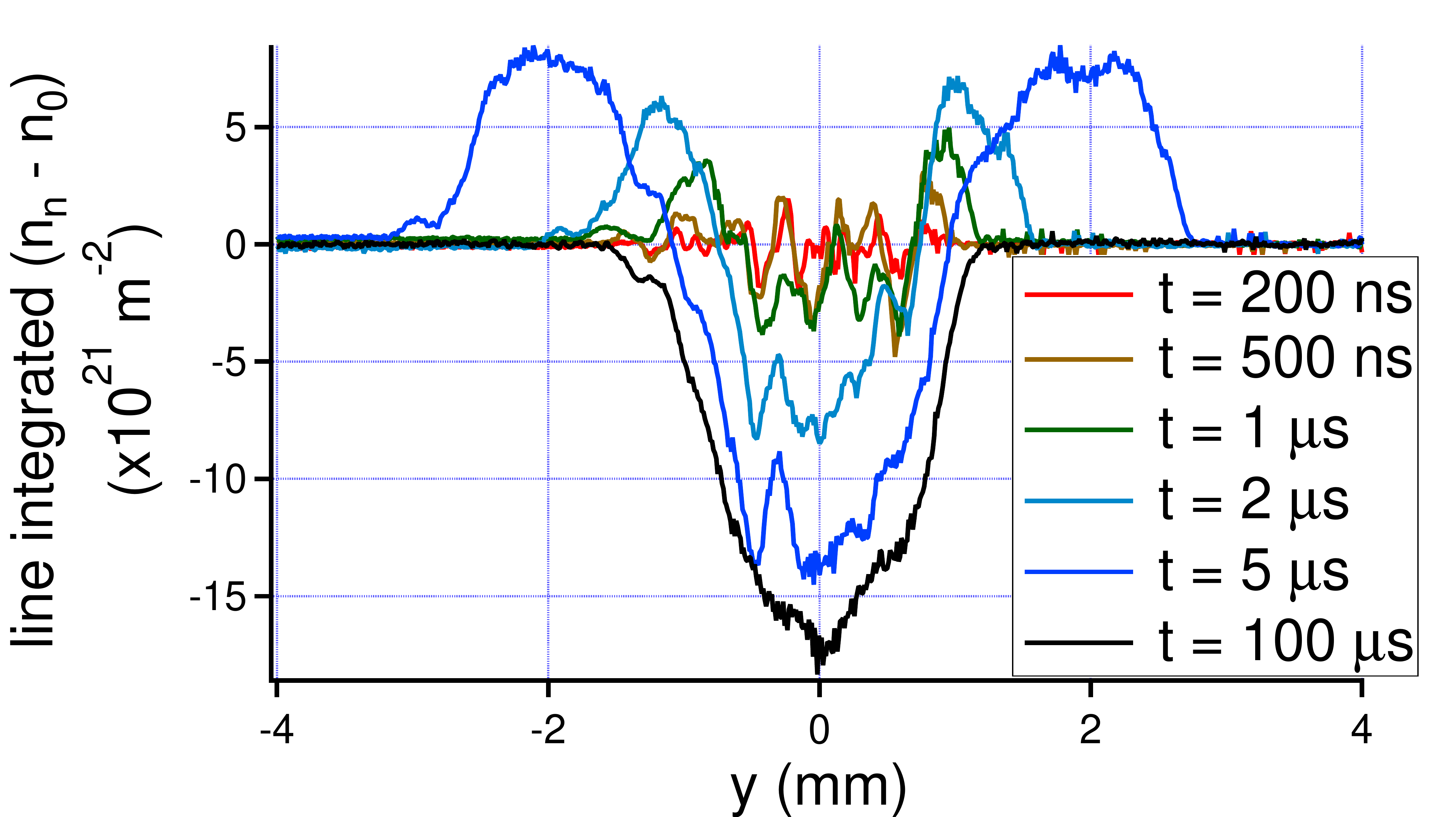}
\end{center}
\caption{Line-integrated air density profiles recorded at $z=\unit{0.96}{\metre}$ after the focusing lens and for different times. Filamentation was generated by a \unit{165}{\milli\joule}, \unit{50}{\femto\second} laser pulse at \unit{800}{\nano\metre} focused at $f/30$.}
\label{figure_5}
\end{figure}

The symmetrization process is highly dependent on the position along the filament bundle and, therefore, on the local profile of energy deposition. For instance, looking at the time evolution of line-integrated air density at position $z = \unit{96}{\centi\metre}$ (figure \ref{figure_5}), we see that in this case, symmetrization takes a much longer time than at $z=\unit{1}{\metre}$ and remains incomplete even after \unit{10}{\micro\second}, but still occurs by \unit{100}{\micro\second}. This gives a clue about the origin of symmetrization. We explain it as follows: multifilamentation results in an inhomogeneous energy deposition, which occurs in several separated channels. Each filament forms a hydrodynamic wave and the associated air underdense channel. If filaments are initially close enough one from the other, as it is the case in the focal zone, individual shock waves quickly interfere constructively, leading to the formation of a single, large shock wave. The time interval during which $u_{SW}$ increases (figure \ref{figure_4}-(c)) can be seen as a sign of this interference process during which the fastest components from the strongest filaments catch up with the slowest components and eventually take the lead. Since in the focal zone at this energy level and with these focusing conditions the laser pulse propagates in the \textit{superfilamentation} regime \cite{Point2014a}, plasma and therefore energy deposition profile are distributed in a much more symmetric fashion than in standard short-scale filamentation areas. This explains why the shock exhibits an almost perfect cylindrical symmetry. As this shock is responsible for the formation of the central underdense channel by ejecting matter from the center, this channel shares the symmetry of the shock wave. When symmetrization is only partial once the shock has left (as displayed in figure \ref{figure_5}), thermal diffusion can act over longer timescales to smooth out any remaining sharp density features, yielding a \textit{diffusive} symmetrization, unlike the pure \textit{acoustic} symmetrization process occurring near the linear focus.

\subsection{Lineic deposited energy}

Still investigating the case of a \unit{165}{\milli\joule}, \unit{50}{\femto\second} laser pulse focused at $f/30$, we estimated the lineic deposited energy using a sonographic technique. It consists in scanning the acoustic emission from the filament bundle along the laser propagation direction by means of a microphone (model 4138 from Br\"uel \& Kjaer) \cite{Yu2003,Point2015a}. Indeed, since heating due to filamentation occurs much more quickly than the characteristic development time of air hydrodynamic response, we can consider it as an isochoric process. If we write the deposited energy as $\Delta U$ then the first law of thermodynamics reads:
\begin{align}
\nonumber\Delta U &= c_vn_0\int_{\mathbb{R}^3}(T(\vv{r})-T_{air})~\mathrm{d}^3\vv{r}\\
&= \frac{c_v}{k_B}\int_{\mathbb{R}^3}(p(\vv{r})-p_{air})~\mathrm{d}^3\vv{r}, 
\end{align}
where $c_v$ is the isochoric heat capacity of an air molecule and $n_0 = \unit{2.47\times10^{25}}{\rpcubic\metre}$ is the air density at $p_{air} = \unit{1.013\times10^5}{\pascal}$ and $T_{air} = \unit{300}{\kelvin}$. In the case of an initial pressure profile with cylindrical symmetry, this equation can be simplified as:
\begin{equation}
\Delta U \approx \frac{\pi c_v}{k_B} \int_{\mathbb{R}}(p_{max}(z)-p_{air}) r_0(z)^2~\mathrm{d}z,
\end{equation}
where $r_0(z)$ is the half width at half maximum (HWHM) of the temperature profile at $z$. Assuming that $r_0$ does not vary significantly along the acoustic source, we find:
\begin{equation}
\Delta U \propto \int_{\mathbb{R}}(p_{max}(z)-p_{air})~\mathrm{d}z.
\label{eq_sonographic_principle}
\end{equation}

Recording the $z$-evolution of the peak pressure with a microphone in the case of an initially cylindrically-symmetric pressure profile then yields the lineic deposited energy. If this method works well in the case of a single filament \cite{Rosenthal2015}, it cannot be directly applied in the present case because this requirement is not fulfilled. The symmetrization process in the focal zone can be used at our advantage. As the system spontaneously evolve to this cylindrically-symmetric state, we can consider it as if it started from a cylindrical pressure profile with a similar initial energy. We then make the assumption that most of energy deposition occurs around the focal zone where this effect is at its strongest, enabling us to retrieve the lineic deposited energy following equation \eqref{eq_sonographic_principle}.


\begin{figure}[!ht]
\begin{center}
\includegraphics[width=.45\textwidth]{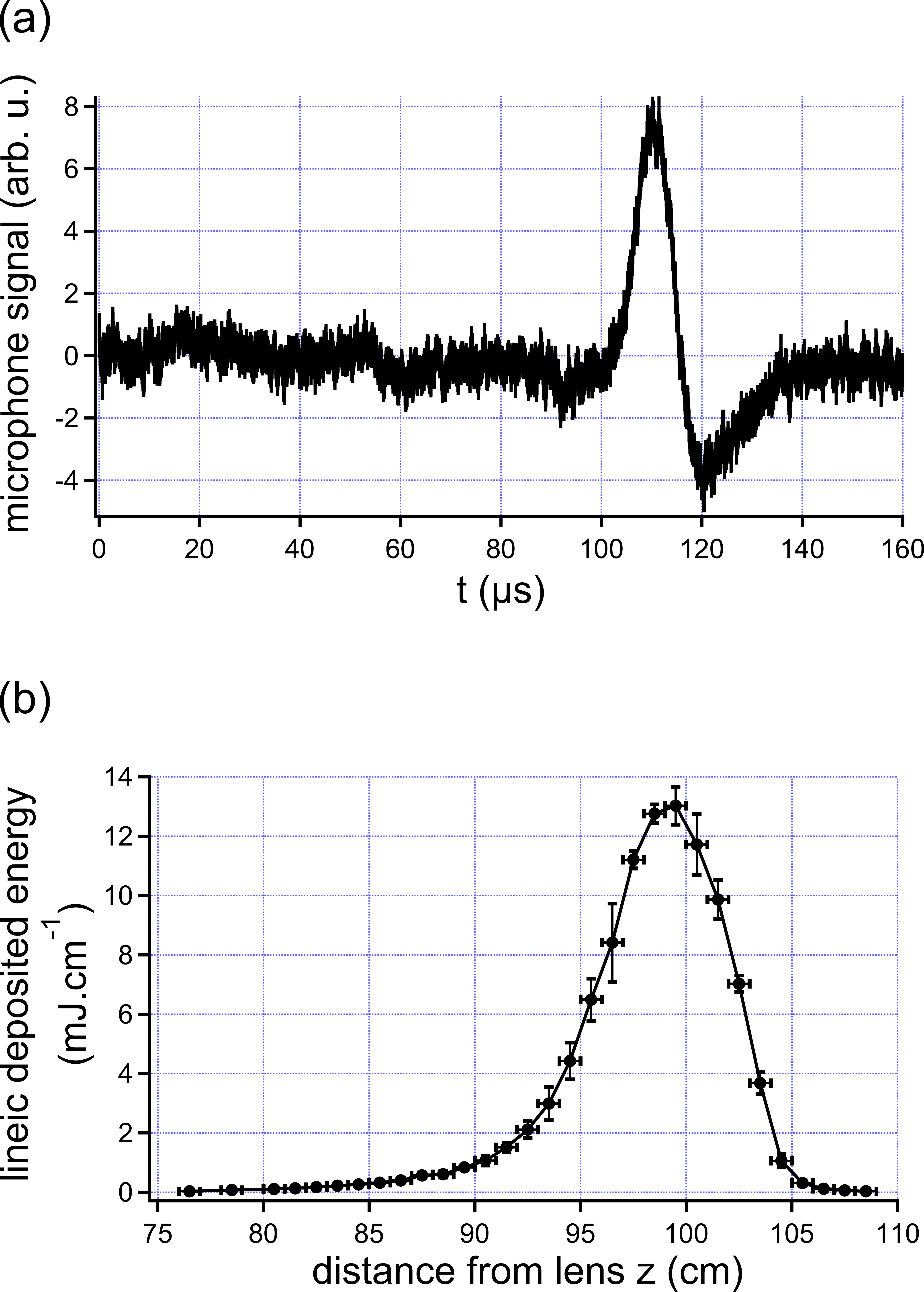}
\end{center}
\caption{(a): example of microphone signal recorded at $z = \unit{80}{\centi\metre}$, displaying only a single acoustic wave. (b): spatial evolution of the lineic deposited energy from a \unit{165}{\milli\joule}, \unit{50}{\femto\second} laser pulse at \unit{800}{\nano\metre} focused at $f/30$. This curve was obtained from sonographic measurements of the bundle acoustic emission and by equating the $z$-integral of the corresponding scan with total deposited energy. Error bars correspond to a confidence interval of one standard deviation evaluated over 200 shots.}
\label{figure_6}
\end{figure}

As seen in figure \ref{figure_6}-(a), the filament bundle already generates a single acoustic wave as far as \unit{20}{\centi\metre} away from the linear focus, showing that at least a partial acoustic symmetrization occurs at this position and strengthening our hypothesis. We therefore performed a full acoustic scan along the filament bundle and equated the integral of the corresponding curve with deposited energy following equation \eqref{eq_sonographic_principle}. Results are plotted in figure \ref{figure_6}-(b). They exhibit a zone about \unit{15}{\centi\metre} long where lineic deposited energy is above \unit{1}{\milli\joule\cdot\centi\rp\metre}, reaching a peak value of \unit{13}{\milli\joule\cdot\centi\rp\metre} shortly before the linear focus, confirming that most of the energy deposition takes place in the focal zone. As a comparison, high-energy single filamentation presented in reference \cite{Point2015} only reached a \unit{400}{\micro\joule\cdot\centi\rp\metre} maximum energy deposition.

\subsection{Physical investigation of energy deposition}

Investigating the underlying physics of laser energy deposition asks for a good knowledge of the different vectors for energy absorption. Plasma generation is the easiest one to characterize. To this purpose we used a spectroscopic analysis of the plasma luminescence. It was shown previously by Th\'eberge \textit{et al.} that emission from the first negative system of the N$_2^+$ cation can be used as a plasma diagnostic following:
\begin{equation}
\int_\mathbb{R}L(z)~\mathrm{d}z \propto N_e,
\end{equation}
where $L$ is the luminescence signal at position $z$ along the multifilament bundle and $N_e$ the total number of free electrons in the plasma \cite{Theberge2006}.

\begin{figure}[!ht]
\begin{center}
\includegraphics[width=.45\textwidth]{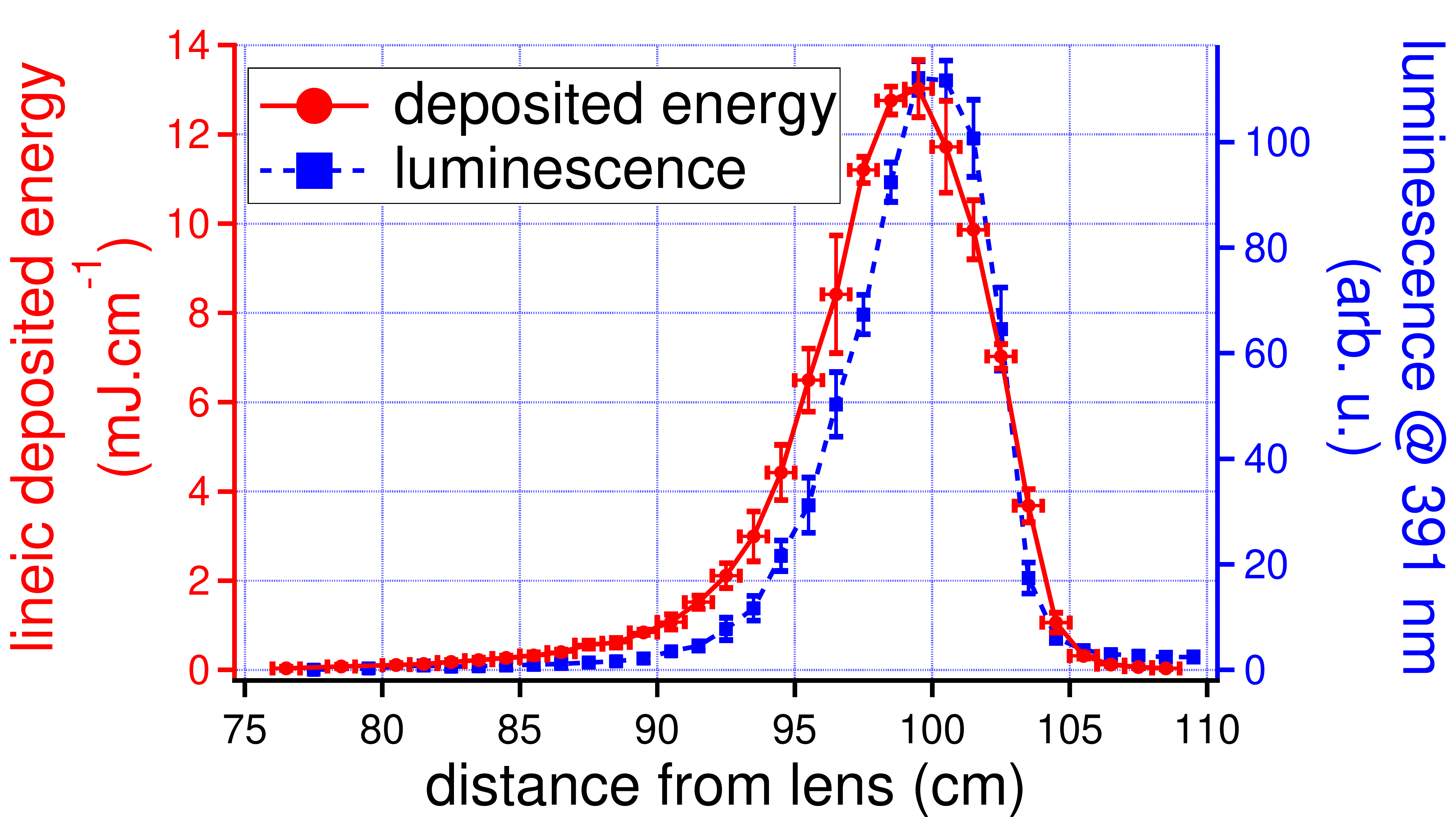}
\end{center}
\caption{(Color online) Plasma luminescence from the first negative system of N$_2^+$ at \unit{391}{\nano\metre} (blue squares) and lineic energy deposition profile (red circles) along the multifilament bundle. Error bars correspond to a confidence interval of one standard deviation evaluated over 200 shots.}
\label{figure_7}
\end{figure}

This means that plasma luminescence from the first negative system of N$_2^+$ is proportional to the area-integrated electron density:
\begin{equation}
L(z) \propto \int_{\mathbb{R}^2}n_e(r,\theta,z)r~\mathrm{d}r\mathrm{d}\theta.
\end{equation}
We recorded plasma luminescence along the multifilament bundle at wavelength \unit{391}{\nano\metre} using a monochromator (model H-20 from Jobin Yvon) with a $\sim \unit{1.5}{\nano\metre}$ spectral resolution. Results are plotted in figure \ref{figure_7}, compared with the lineic deposited energy profile.

From this figure, one can see that the total plasma length is in excess of \unit{15}{\centi\metre}, starting slightly before $z = \unit{90}{\centi\metre}$ and ending between $z = 105$ and $z = \unit{110}{\centi\metre}$. It is interesting to note that energy deposition clearly starts increasing before plasma luminescence, indicating that in the area before the focus ionization alone is not the main channel for energy deposition. We attribute this behavior to the occurrence of stimulated rotational Raman scattering on air molecules \cite{Seideman2001}. Conversely, once in the focal zone, lineic deposited energy and plasma luminescence behave quite similarly, decreasing simultaneously and almost completely disappearing past $z = \unit{105}{\centi\metre}$. This good agreement shows ionization becomes the main channel for energy deposition close to the focus.

Assuming that the peak energy deposition is reached mostly due to ionization of O$_2$ molecules \cite{Couairon2007}, the peak lineic deposited energy is given by:
\begin{align}
\nonumber\left.\frac{\partial\Delta U}{\partial z}\right|_{max} &= \int_{\mathbb{R}^2}n_{e,max}(r,z_{max})U_{\mathrm{O}_2}r~\mathrm{d}r\mathrm{d}\theta\\
&\approx n_{e,max}(z_{max})U_{\mathrm{O}_2}\pi r_0^2,
\end{align}
where $U_{\mathrm{O}_2} \approx \unit{12}{\electronvolt}$ is the ionization potential of O$_2$, $z_{max}$ the position along the multifilament bundle where the peak lineic energy deposition is reached and $r_0$ the HWHM of the electron density profile. Peak electron density in the case of single filaments generated using a $\sim f/30$ focusing has been recorded \cite{Theberge2006} and numerically estimated \cite{Kiran2010a} to be on the order of $\unit{3\times10^{23}}{\rpcubic\metre}$. Since \textit{superfilamentation} typically results in a peak electron density one order of magnitude higher than in equivalent single filaments, with an equivalent radius in the focal zone of $\sim\unit{300}{\micro\metre}$ \cite{Point2014a}, the peak lineic deposited energy is therefore estimated at:
\begin{equation}
\left.\frac{\partial\Delta U}{\partial z}\right|_{max} \sim \unit{15}{\milli\joule\cdot\centi\rp\metre},
\end{equation}
which is indeed close to the measured peak value. 

Energy deposition then seems mainly limited by two different factors: first, electron density that can be reached during pulse propagation and, second, the effective length of the plasma column. Given the results presented in this Article, it is clear that increasing the focusing leads to a rise of peak electron density, compounded by a slower shrinking of filamentation length, leading to an overall increase of energy deposition. However a too strong focusing is detrimental to energy deposition because of the generation of a very localized and strong plasma, preventing the pulse energy from being efficiently focused and deposited in air.

\section{Conclusion}

In this Article, we presented results from the study of energy deposition in air in the high energy, multifilamentation regime. Influence of experimental parameters on total energy deposition was first investigated. It was found that deposited energy increases with the laser pulse peak power, both in terms of energy and duration. Influence of pulse duration is important, with slight deviations from the minimum duration yielding a significantly lower energy deposition. Influence of pulse energy depends on whether the beam forms a single or multifilaments. In the first case, energy deposition quickly increases with input energy as more and more energy is channeled through the single structure. In the second case, deposited energy still increases with input energy, but does so at a far lower rate as excess energy is redistributed over many competing filaments. As for focusing conditions, we found that strong focusing yielded a significantly higher energy deposition than weak focusing. However increasing the focusing too much actually leads to a significantly lower energy deposition, due to the strong generated plasma, preventing from efficiently channeling laser energy at the focal point. Optimal focusing was evaluated at $f/30$ for a \unit{165}{\milli\joule}  pulse energy and a \unit{50}{\femto\second} pulse duration, yielding the deposition of more than 60 \% of the input energy.

Filamentation-induced hydrodynamics were then studied by means of transverse interferometry. As the beam propagates in the multifilamentation regime, initial energy deposition takes place in many parallel structures in an inhomogeneous fashion. However in the focal zone, where \textit{superfilaments} occur \cite{Point2014a}, air structures spontaneously evolve towards a quasi cylindrically symmetrical state after a delay typically on the order of \unit{1}{\micro\second}. We explain this phenomenon by the constructive interference between individual shock waves generated by each superfilament, and the relatively cylindrical distribution of these superfilaments. Going away from the focal zone, symmetrization due to the shock wave process is only partial but still happens over longer timescales ($\sim \unit{100}{\micro\second}$) due to thermal diffusion. This symmetrization process enabled us to use a sonographic analysis to estimate the spatial evolution of the lineic deposited energy. It was found that energy densities above \unit{1}{\milli\joule\cdot\centi\rp\metre} could be reach over \unit{15}{\centi\metre}, with a peak lineic energy of more than \unit{1}{\joule\cdot\rp\metre}.

These results are particularly interesting in the prospect of filamentation-induced air virtual optical structures, such as virtual waveguides. Multifilamentation indeed gives birth to significantly deeper and longer-lived underdense channel that could be used in a similar way as described in reference \cite{Jhajj2014}. A major drawback of multifilamentation comes from the inhomogeneity of the whole beam, especially the existence of a zone where no symmetrization occurs. Still, some degree of control can be achieved over filamentation occurrence \cite{Mechain2004}, which could help shape energy deposition profile. Also, using a smoother and cleaner beam spatial profile would decrease the modulational instability of the beam and allow to have a higher threshold for the appearance of multifilaments \cite{Fibich2005}.

\bibliographystyle{apsrev4-1}
\bibliography{biblio}

\begin{thebibliography}{25}%
\makeatletter
\providecommand \@ifxundefined [1]{%
 \@ifx{#1\undefined}
}%
\providecommand \@ifnum [1]{%
 \ifnum #1\expandafter \@firstoftwo
 \else \expandafter \@secondoftwo
 \fi
}%
\providecommand \@ifx [1]{%
 \ifx #1\expandafter \@firstoftwo
 \else \expandafter \@secondoftwo
 \fi
}%
\providecommand \natexlab [1]{#1}%
\providecommand \enquote  [1]{``#1''}%
\providecommand \bibnamefont  [1]{#1}%
\providecommand \bibfnamefont [1]{#1}%
\providecommand \citenamefont [1]{#1}%
\providecommand \href@noop [0]{\@secondoftwo}%
\providecommand \href [0]{\begingroup \@sanitize@url \@href}%
\providecommand \@href[1]{\@@startlink{#1}\@@href}%
\providecommand \@@href[1]{\endgroup#1\@@endlink}%
\providecommand \@sanitize@url [0]{\catcode `\\12\catcode `\$12\catcode
  `\&12\catcode `\#12\catcode `\^12\catcode `\_12\catcode `\%12\relax}%
\providecommand \@@startlink[1]{}%
\providecommand \@@endlink[0]{}%
\providecommand \url  [0]{\begingroup\@sanitize@url \@url }%
\providecommand \@url [1]{\endgroup\@href {#1}{\urlprefix }}%
\providecommand \urlprefix  [0]{URL }%
\providecommand \Eprint [0]{\href }%
\providecommand \doibase [0]{http://dx.doi.org/}%
\providecommand \selectlanguage [0]{\@gobble}%
\providecommand \bibinfo  [0]{\@secondoftwo}%
\providecommand \bibfield  [0]{\@secondoftwo}%
\providecommand \translation [1]{[#1]}%
\providecommand \BibitemOpen [0]{}%
\providecommand \bibitemStop [0]{}%
\providecommand \bibitemNoStop [0]{.\EOS\space}%
\providecommand \EOS [0]{\spacefactor3000\relax}%
\providecommand \BibitemShut  [1]{\csname bibitem#1\endcsname}%
\let\auto@bib@innerbib\@empty
\bibitem [{\citenamefont {Couairon}\ and\ \citenamefont
  {Mysyrowicz}(2007)}]{Couairon2007}%
  \BibitemOpen
  \bibfield  {author} {\bibinfo {author} {\bibfnamefont {A.}~\bibnamefont
  {Couairon}}\ and\ \bibinfo {author} {\bibfnamefont {A.}~\bibnamefont
  {Mysyrowicz}},\ }\href {http://dx.doi.org/10.1016/j.physrep.2006.12.005}
  {\bibfield  {journal} {\bibinfo  {journal} {Physics Reports}\ }\textbf
  {\bibinfo {volume} {441}},\ \bibinfo {pages} {47} (\bibinfo {year}
  {2007})}\BibitemShut {NoStop}%
\bibitem [{\citenamefont {Nibbering}\ \emph {et~al.}(1997)\citenamefont
  {Nibbering}, \citenamefont {Grillon}, \citenamefont {Franco}, \citenamefont
  {Prade},\ and\ \citenamefont {Mysyrowicz}}]{Nibbering1997}%
  \BibitemOpen
  \bibfield  {author} {\bibinfo {author} {\bibfnamefont {E.~T.~J.}\
  \bibnamefont {Nibbering}}, \bibinfo {author} {\bibfnamefont {G.}~\bibnamefont
  {Grillon}}, \bibinfo {author} {\bibfnamefont {M.~A.}\ \bibnamefont {Franco}},
  \bibinfo {author} {\bibfnamefont {B.~S.}\ \bibnamefont {Prade}}, \ and\
  \bibinfo {author} {\bibfnamefont {A.}~\bibnamefont {Mysyrowicz}},\ }\href
  {http://dx.doi.org/10.1364/JOSAB.14.000650} {\bibfield  {journal} {\bibinfo
  {journal} {Journal of the Optical Society of America B}\ }\textbf {\bibinfo
  {volume} {14}},\ \bibinfo {pages} {650} (\bibinfo {year} {1997})}\BibitemShut
  {NoStop}%
\bibitem [{\citenamefont {Seideman}(2001)}]{Seideman2001}%
  \BibitemOpen
  \bibfield  {author} {\bibinfo {author} {\bibfnamefont {T.}~\bibnamefont
  {Seideman}},\ }\href {http://dx.doi.org/10.1063/1.1400131} {\bibfield
  {journal} {\bibinfo  {journal} {The Journal of Chemical Physics}\ }\textbf
  {\bibinfo {volume} {115}},\ \bibinfo {pages} {5965} (\bibinfo {year}
  {2001})}\BibitemShut {NoStop}%
\bibitem [{\citenamefont {Kartashov}\ \emph {et~al.}(2006)\citenamefont
  {Kartashov}, \citenamefont {Kirsanov}, \citenamefont {Kiselev}, \citenamefont
  {Stepanov}, \citenamefont {Bochkarev}, \citenamefont {Ponomarev},\ and\
  \citenamefont {Tikhomirov}}]{Kartashov2006}%
  \BibitemOpen
  \bibfield  {author} {\bibinfo {author} {\bibfnamefont {D.~V.}\ \bibnamefont
  {Kartashov}}, \bibinfo {author} {\bibfnamefont {A.~V.}\ \bibnamefont
  {Kirsanov}}, \bibinfo {author} {\bibfnamefont {A.~M.}\ \bibnamefont
  {Kiselev}}, \bibinfo {author} {\bibfnamefont {A.~N.}\ \bibnamefont
  {Stepanov}}, \bibinfo {author} {\bibfnamefont {N.~N.}\ \bibnamefont
  {Bochkarev}}, \bibinfo {author} {\bibfnamefont {Y.~N.}\ \bibnamefont
  {Ponomarev}}, \ and\ \bibinfo {author} {\bibfnamefont {B.~A.}\ \bibnamefont
  {Tikhomirov}},\ }\href {\doibase 10.1364/OE.14.007552} {\bibfield  {journal}
  {\bibinfo  {journal} {Optics Express}\ }\textbf {\bibinfo {volume} {14}},\
  \bibinfo {pages} {7552} (\bibinfo {year} {2006})}\BibitemShut {NoStop}%
\bibitem [{\citenamefont {Tzortzakis}\ \emph {et~al.}(2000)\citenamefont
  {Tzortzakis}, \citenamefont {Prade}, \citenamefont {Franco},\ and\
  \citenamefont {Mysyrowicz}}]{Tzortzakis2000}%
  \BibitemOpen
  \bibfield  {author} {\bibinfo {author} {\bibfnamefont {S.}~\bibnamefont
  {Tzortzakis}}, \bibinfo {author} {\bibfnamefont {B.}~\bibnamefont {Prade}},
  \bibinfo {author} {\bibfnamefont {M.}~\bibnamefont {Franco}}, \ and\ \bibinfo
  {author} {\bibfnamefont {A.}~\bibnamefont {Mysyrowicz}},\ }\href {\doibase
  10.1016/S0030-4018(00)00734-3} {\bibfield  {journal} {\bibinfo  {journal}
  {Optics Communications}\ }\textbf {\bibinfo {volume} {181}},\ \bibinfo
  {pages} {123} (\bibinfo {year} {2000})}\BibitemShut {NoStop}%
\bibitem [{\citenamefont {Chen}\ \emph {et~al.}(2007)\citenamefont {Chen},
  \citenamefont {Varma}, \citenamefont {York},\ and\ \citenamefont
  {Milchberg}}]{Chen2007}%
  \BibitemOpen
  \bibfield  {author} {\bibinfo {author} {\bibfnamefont {Y.-H.}\ \bibnamefont
  {Chen}}, \bibinfo {author} {\bibfnamefont {S.}~\bibnamefont {Varma}},
  \bibinfo {author} {\bibfnamefont {A.}~\bibnamefont {York}}, \ and\ \bibinfo
  {author} {\bibfnamefont {H.~M.}\ \bibnamefont {Milchberg}},\ }\href {\doibase
  10.1364/OE.15.011341} {\bibfield  {journal} {\bibinfo  {journal} {Optics
  Express}\ }\textbf {\bibinfo {volume} {15}},\ \bibinfo {pages} {11341}
  (\bibinfo {year} {2007})}\BibitemShut {NoStop}%
\bibitem [{\citenamefont {Cheng}\ \emph {et~al.}(2013)\citenamefont {Cheng},
  \citenamefont {Wahlstrand}, \citenamefont {Jhajj},\ and\ \citenamefont
  {Milchberg}}]{Cheng2013}%
  \BibitemOpen
  \bibfield  {author} {\bibinfo {author} {\bibfnamefont {Y.-H.}\ \bibnamefont
  {Cheng}}, \bibinfo {author} {\bibfnamefont {J.~K.}\ \bibnamefont
  {Wahlstrand}}, \bibinfo {author} {\bibfnamefont {N.}~\bibnamefont {Jhajj}}, \
  and\ \bibinfo {author} {\bibfnamefont {H.~M.}\ \bibnamefont {Milchberg}},\
  }\href {http://dx.doi.org/10.1364/oe.21.004740} {\bibfield  {journal}
  {\bibinfo  {journal} {Optics Express}\ }\textbf {\bibinfo {volume} {21}},\
  \bibinfo {pages} {4740} (\bibinfo {year} {2013})}\BibitemShut {NoStop}%
\bibitem [{\citenamefont {Point}\ \emph {et~al.}(2015)\citenamefont {Point},
  \citenamefont {Mili\'an}, \citenamefont {Couairon}, \citenamefont
  {Mysyrowicz},\ and\ \citenamefont {Houard}}]{Point2015}%
  \BibitemOpen
  \bibfield  {author} {\bibinfo {author} {\bibfnamefont {G.}~\bibnamefont
  {Point}}, \bibinfo {author} {\bibfnamefont {C.}~\bibnamefont {Mili\'an}},
  \bibinfo {author} {\bibfnamefont {A.}~\bibnamefont {Couairon}}, \bibinfo
  {author} {\bibfnamefont {A.}~\bibnamefont {Mysyrowicz}}, \ and\ \bibinfo
  {author} {\bibfnamefont {A.}~\bibnamefont {Houard}},\ }\href
  {http://dx.doi.org/10.1088/0953-4075/48/9/094009} {\bibfield  {journal}
  {\bibinfo  {journal} {Journal of Physics B: Atomic, Molecular and Optical
  Physics}\ }\textbf {\bibinfo {volume} {48}},\ \bibinfo {pages} {094009}
  (\bibinfo {year} {2015})}\BibitemShut {NoStop}%
\bibitem [{\citenamefont {Jhajj}\ \emph {et~al.}(2014)\citenamefont {Jhajj},
  \citenamefont {Rosenthal}, \citenamefont {Birnbaum}, \citenamefont
  {Wahlstrand},\ and\ \citenamefont {Milchberg}}]{Jhajj2014}%
  \BibitemOpen
  \bibfield  {author} {\bibinfo {author} {\bibfnamefont {N.}~\bibnamefont
  {Jhajj}}, \bibinfo {author} {\bibfnamefont {E.~W.}\ \bibnamefont
  {Rosenthal}}, \bibinfo {author} {\bibfnamefont {R.}~\bibnamefont {Birnbaum}},
  \bibinfo {author} {\bibfnamefont {J.~K.}\ \bibnamefont {Wahlstrand}}, \ and\
  \bibinfo {author} {\bibfnamefont {H.~M.}\ \bibnamefont {Milchberg}},\ }\href
  {\doibase 10.1103/PhysRevX.4.011027} {\bibfield  {journal} {\bibinfo
  {journal} {Physical Review X}\ }\textbf {\bibinfo {volume} {4}},\ \bibinfo
  {pages} {011027} (\bibinfo {year} {2014})}\BibitemShut {NoStop}%
\bibitem [{\citenamefont {Lahav}\ \emph {et~al.}(2014)\citenamefont {Lahav},
  \citenamefont {Levi}, \citenamefont {Orr}, \citenamefont {Nemirovsky},
  \citenamefont {Nemirovsky}, \citenamefont {Kaminer}, \citenamefont {Segev},\
  and\ \citenamefont {Cohen}}]{Lahav2014}%
  \BibitemOpen
  \bibfield  {author} {\bibinfo {author} {\bibfnamefont {O.}~\bibnamefont
  {Lahav}}, \bibinfo {author} {\bibfnamefont {L.}~\bibnamefont {Levi}},
  \bibinfo {author} {\bibfnamefont {I.}~\bibnamefont {Orr}}, \bibinfo {author}
  {\bibfnamefont {R.~A.}\ \bibnamefont {Nemirovsky}}, \bibinfo {author}
  {\bibfnamefont {J.}~\bibnamefont {Nemirovsky}}, \bibinfo {author}
  {\bibfnamefont {I.}~\bibnamefont {Kaminer}}, \bibinfo {author} {\bibfnamefont
  {M.}~\bibnamefont {Segev}}, \ and\ \bibinfo {author} {\bibfnamefont
  {O.}~\bibnamefont {Cohen}},\ }\href {\doibase 10.1103/PhysRevA.90.021801}
  {\bibfield  {journal} {\bibinfo  {journal} {Physical Review A}\ }\textbf
  {\bibinfo {volume} {90}},\ \bibinfo {pages} {021801} (\bibinfo {year}
  {2014})}\BibitemShut {NoStop}%
\bibitem [{\citenamefont {Dufour}\ \emph {et~al.}(2013)\citenamefont {Dufour},
  \citenamefont {Fornet},\ and\ \citenamefont {Rogier}}]{Dufour2013}%
  \BibitemOpen
  \bibfield  {author} {\bibinfo {author} {\bibfnamefont {G.}~\bibnamefont
  {Dufour}}, \bibinfo {author} {\bibfnamefont {B.}~\bibnamefont {Fornet}}, \
  and\ \bibinfo {author} {\bibfnamefont {F.}~\bibnamefont {Rogier}},\ }\href
  {http://dx.doi.org/10.1504/IJAD.2013.050924} {\bibfield  {journal} {\bibinfo
  {journal} {International Journal of Aerodynamics}\ }\textbf {\bibinfo
  {volume} {3}},\ \bibinfo {pages} {122} (\bibinfo {year} {2013})}\BibitemShut
  {NoStop}%
\bibitem [{\citenamefont {Mlejnek}\ \emph {et~al.}(1999)\citenamefont
  {Mlejnek}, \citenamefont {Kolesik}, \citenamefont {Moloney},\ and\
  \citenamefont {Wright}}]{Mlejnek1999}%
  \BibitemOpen
  \bibfield  {author} {\bibinfo {author} {\bibfnamefont {M.}~\bibnamefont
  {Mlejnek}}, \bibinfo {author} {\bibfnamefont {M.}~\bibnamefont {Kolesik}},
  \bibinfo {author} {\bibfnamefont {J.~V.}\ \bibnamefont {Moloney}}, \ and\
  \bibinfo {author} {\bibfnamefont {E.~M.}\ \bibnamefont {Wright}},\ }\href
  {http://dx.doi.org/10.1103/PhysRevLett.83.2938} {\bibfield  {journal}
  {\bibinfo  {journal} {Physical Review Letters}\ }\textbf {\bibinfo {volume}
  {83}},\ \bibinfo {pages} {2938} (\bibinfo {year} {1999})}\BibitemShut
  {NoStop}%
\bibitem [{\citenamefont {Fibich}\ \emph {et~al.}(2005)\citenamefont {Fibich},
  \citenamefont {Eisenmann}, \citenamefont {Ilan}, \citenamefont {Erlich},
  \citenamefont {Fraenkel}, \citenamefont {Henis}, \citenamefont {Gaeta},\ and\
  \citenamefont {Zigler}}]{Fibich2005}%
  \BibitemOpen
  \bibfield  {author} {\bibinfo {author} {\bibfnamefont {G.}~\bibnamefont
  {Fibich}}, \bibinfo {author} {\bibfnamefont {S.}~\bibnamefont {Eisenmann}},
  \bibinfo {author} {\bibfnamefont {B.}~\bibnamefont {Ilan}}, \bibinfo {author}
  {\bibfnamefont {Y.}~\bibnamefont {Erlich}}, \bibinfo {author} {\bibfnamefont
  {M.}~\bibnamefont {Fraenkel}}, \bibinfo {author} {\bibfnamefont
  {Z.}~\bibnamefont {Henis}}, \bibinfo {author} {\bibfnamefont
  {A.}~\bibnamefont {Gaeta}}, \ and\ \bibinfo {author} {\bibfnamefont
  {A.}~\bibnamefont {Zigler}},\ }\href
  {http://dx.doi.org/10.1364/OPEX.13.005897} {\bibfield  {journal} {\bibinfo
  {journal} {Optics Express}\ }\textbf {\bibinfo {volume} {13}},\ \bibinfo
  {pages} {5897} (\bibinfo {year} {2005})}\BibitemShut {NoStop}%
\bibitem [{\citenamefont {Yu}\ \emph {et~al.}(2003)\citenamefont {Yu},
  \citenamefont {Mondelain}, \citenamefont {Kasparian}, \citenamefont {Salmon},
  \citenamefont {Geffroy}, \citenamefont {Favre}, \citenamefont {Boutou},\ and\
  \citenamefont {Wolf}}]{Yu2003}%
  \BibitemOpen
  \bibfield  {author} {\bibinfo {author} {\bibfnamefont {J.}~\bibnamefont
  {Yu}}, \bibinfo {author} {\bibfnamefont {D.}~\bibnamefont {Mondelain}},
  \bibinfo {author} {\bibfnamefont {J.}~\bibnamefont {Kasparian}}, \bibinfo
  {author} {\bibfnamefont {E.}~\bibnamefont {Salmon}}, \bibinfo {author}
  {\bibfnamefont {S.}~\bibnamefont {Geffroy}}, \bibinfo {author} {\bibfnamefont
  {C.}~\bibnamefont {Favre}}, \bibinfo {author} {\bibfnamefont
  {V.}~\bibnamefont {Boutou}}, \ and\ \bibinfo {author} {\bibfnamefont {J.-P.}\
  \bibnamefont {Wolf}},\ }\href {\doibase 10.1364/AO.42.007117} {\bibfield
  {journal} {\bibinfo  {journal} {Applied Optics}\ }\textbf {\bibinfo {volume}
  {42}},\ \bibinfo {pages} {7117} (\bibinfo {year} {2003})}\BibitemShut
  {NoStop}%
\bibitem [{\citenamefont {Point}(2015)}]{Point2015a}%
  \BibitemOpen
  \bibfield  {author} {\bibinfo {author} {\bibfnamefont {G.}~\bibnamefont
  {Point}},\ }\emph {\bibinfo {title} {Energy deposition in air from
  femtosecond laser filamentation for the control of high voltage spark
  discharges}},\ \href {https://tel.archives-ouvertes.fr/tel-01202982} {Ph.D.
  thesis},\ \bibinfo  {school} {Ecole Polytechnique, France} (\bibinfo {year}
  {2015})\BibitemShut {NoStop}%
\bibitem [{\citenamefont {Raizer}(1966)}]{Raizer1966}%
  \BibitemOpen
  \bibfield  {author} {\bibinfo {author} {\bibfnamefont {Y.~P.}\ \bibnamefont
  {Raizer}},\ }\href {http://dx.doi.org/10.1070/PU1966v008n05ABEH003027}
  {\bibfield  {journal} {\bibinfo  {journal} {Soviet Physics Uspekhi}\ }\textbf
  {\bibinfo {volume} {8}},\ \bibinfo {pages} {650} (\bibinfo {year}
  {1966})}\BibitemShut {NoStop}%
\bibitem [{\citenamefont {Mermillod-Blondin}\ \emph {et~al.}(2008)\citenamefont
  {Mermillod-Blondin}, \citenamefont {Mauclair}, \citenamefont {Rosenfeld},
  \citenamefont {Bonse}, \citenamefont {Hertel}, \citenamefont {Audouard},\
  and\ \citenamefont {Stoian}}]{Mermillod2008}%
  \BibitemOpen
  \bibfield  {author} {\bibinfo {author} {\bibfnamefont {A.}~\bibnamefont
  {Mermillod-Blondin}}, \bibinfo {author} {\bibfnamefont {C.}~\bibnamefont
  {Mauclair}}, \bibinfo {author} {\bibfnamefont {A.}~\bibnamefont {Rosenfeld}},
  \bibinfo {author} {\bibfnamefont {J.}~\bibnamefont {Bonse}}, \bibinfo
  {author} {\bibfnamefont {I.~V.}\ \bibnamefont {Hertel}}, \bibinfo {author}
  {\bibfnamefont {E.}~\bibnamefont {Audouard}}, \ and\ \bibinfo {author}
  {\bibfnamefont {R.}~\bibnamefont {Stoian}},\ }\href
  {http://dx.doi.org/10.1063/1.2958345} {\bibfield  {journal} {\bibinfo
  {journal} {Applied Physics Letters}\ }\textbf {\bibinfo {volume} {93}},\
  \bibinfo {pages} {021921} (\bibinfo {year} {2008})}\BibitemShut {NoStop}%
\bibitem [{\citenamefont {Brelet}\ \emph {et~al.}(2015)\citenamefont {Brelet},
  \citenamefont {Jarnac}, \citenamefont {Carbonnel}, \citenamefont {Andr\'e},
  \citenamefont {Mysyrowicz}, \citenamefont {Houard}, \citenamefont
  {Fattaccioli}, \citenamefont {Guillermin},\ and\ \citenamefont
  {Sessarego}}]{Brelet2015}%
  \BibitemOpen
  \bibfield  {author} {\bibinfo {author} {\bibfnamefont {Y.}~\bibnamefont
  {Brelet}}, \bibinfo {author} {\bibfnamefont {A.}~\bibnamefont {Jarnac}},
  \bibinfo {author} {\bibfnamefont {J.}~\bibnamefont {Carbonnel}}, \bibinfo
  {author} {\bibfnamefont {Y.-B.}\ \bibnamefont {Andr\'e}}, \bibinfo {author}
  {\bibfnamefont {A.}~\bibnamefont {Mysyrowicz}}, \bibinfo {author}
  {\bibfnamefont {A.}~\bibnamefont {Houard}}, \bibinfo {author} {\bibfnamefont
  {D.}~\bibnamefont {Fattaccioli}}, \bibinfo {author} {\bibfnamefont
  {R.}~\bibnamefont {Guillermin}}, \ and\ \bibinfo {author} {\bibfnamefont
  {J.-P.}\ \bibnamefont {Sessarego}},\ }\href
  {http://dx.doi.org/10.1121/1.4914998} {\bibfield  {journal} {\bibinfo
  {journal} {The Journal of the Acoustical Society of America}\ }\textbf
  {\bibinfo {volume} {137}},\ \bibinfo {pages} {288} (\bibinfo {year}
  {2015})}\BibitemShut {NoStop}%
\bibitem [{\citenamefont {Kiran}\ \emph
  {et~al.}(2010{\natexlab{a}})\citenamefont {Kiran}, \citenamefont {Bagchi},
  \citenamefont {Arnold}, \citenamefont {Krishnan}, \citenamefont {Kumar},\
  and\ \citenamefont {Couairon}}]{Kiran2010}%
  \BibitemOpen
  \bibfield  {author} {\bibinfo {author} {\bibfnamefont {P.~P.}\ \bibnamefont
  {Kiran}}, \bibinfo {author} {\bibfnamefont {S.}~\bibnamefont {Bagchi}},
  \bibinfo {author} {\bibfnamefont {C.~L.}\ \bibnamefont {Arnold}}, \bibinfo
  {author} {\bibfnamefont {S.~R.}\ \bibnamefont {Krishnan}}, \bibinfo {author}
  {\bibfnamefont {G.~R.}\ \bibnamefont {Kumar}}, \ and\ \bibinfo {author}
  {\bibfnamefont {A.}~\bibnamefont {Couairon}},\ }\href {\doibase
  10.1364/OE.18.021504} {\bibfield  {journal} {\bibinfo  {journal} {Optics
  Express}\ }\textbf {\bibinfo {volume} {18}},\ \bibinfo {pages} {21504}
  (\bibinfo {year} {2010}{\natexlab{a}})}\BibitemShut {NoStop}%
\bibitem [{\citenamefont {Point}\ \emph
  {et~al.}(2014{\natexlab{a}})\citenamefont {Point}, \citenamefont {Brelet},
  \citenamefont {Arantchouk}, \citenamefont {Carbonnel}, \citenamefont {Prade},
  \citenamefont {Mysyrowicz},\ and\ \citenamefont {Houard}}]{Point2014}%
  \BibitemOpen
  \bibfield  {author} {\bibinfo {author} {\bibfnamefont {G.}~\bibnamefont
  {Point}}, \bibinfo {author} {\bibfnamefont {Y.}~\bibnamefont {Brelet}},
  \bibinfo {author} {\bibfnamefont {L.}~\bibnamefont {Arantchouk}}, \bibinfo
  {author} {\bibfnamefont {J.}~\bibnamefont {Carbonnel}}, \bibinfo {author}
  {\bibfnamefont {B.}~\bibnamefont {Prade}}, \bibinfo {author} {\bibfnamefont
  {A.}~\bibnamefont {Mysyrowicz}}, \ and\ \bibinfo {author} {\bibfnamefont
  {A.}~\bibnamefont {Houard}},\ }\href {http://dx.doi.org/10.1063/1.4902533}
  {\bibfield  {journal} {\bibinfo  {journal} {Review of Scientific
  Instruments}\ }\textbf {\bibinfo {volume} {85}},\ \bibinfo {pages} {123101}
  (\bibinfo {year} {2014}{\natexlab{a}})}\BibitemShut {NoStop}%
\bibitem [{\citenamefont {Point}\ \emph
  {et~al.}(2014{\natexlab{b}})\citenamefont {Point}, \citenamefont {Brelet},
  \citenamefont {Houard}, \citenamefont {Jukna}, \citenamefont {Mili\'an},
  \citenamefont {Carbonnel}, \citenamefont {Liu}, \citenamefont {Couairon},\
  and\ \citenamefont {Mysyrowicz}}]{Point2014a}%
  \BibitemOpen
  \bibfield  {author} {\bibinfo {author} {\bibfnamefont {G.}~\bibnamefont
  {Point}}, \bibinfo {author} {\bibfnamefont {Y.}~\bibnamefont {Brelet}},
  \bibinfo {author} {\bibfnamefont {A.}~\bibnamefont {Houard}}, \bibinfo
  {author} {\bibfnamefont {V.}~\bibnamefont {Jukna}}, \bibinfo {author}
  {\bibfnamefont {C.}~\bibnamefont {Mili\'an}}, \bibinfo {author}
  {\bibfnamefont {J.}~\bibnamefont {Carbonnel}}, \bibinfo {author}
  {\bibfnamefont {Y.}~\bibnamefont {Liu}}, \bibinfo {author} {\bibfnamefont
  {A.}~\bibnamefont {Couairon}}, \ and\ \bibinfo {author} {\bibfnamefont
  {A.}~\bibnamefont {Mysyrowicz}},\ }\href {\doibase
  10.1103/PhysRevLett.112.223902} {\bibfield  {journal} {\bibinfo  {journal}
  {Physical Review Letters}\ }\textbf {\bibinfo {volume} {112}},\ \bibinfo
  {pages} {223902} (\bibinfo {year} {2014}{\natexlab{b}})}\BibitemShut
  {NoStop}%
\bibitem [{\citenamefont {Rosenthal}\ \emph {et~al.}(2015)\citenamefont
  {Rosenthal}, \citenamefont {Palastro}, \citenamefont {Jhajj}, \citenamefont
  {Zahedpour}, \citenamefont {Wahlstrand},\ and\ \citenamefont
  {Milchberg}}]{Rosenthal2015}%
  \BibitemOpen
  \bibfield  {author} {\bibinfo {author} {\bibfnamefont {E.~W.}\ \bibnamefont
  {Rosenthal}}, \bibinfo {author} {\bibfnamefont {J.~P.}\ \bibnamefont
  {Palastro}}, \bibinfo {author} {\bibfnamefont {N.}~\bibnamefont {Jhajj}},
  \bibinfo {author} {\bibfnamefont {S.}~\bibnamefont {Zahedpour}}, \bibinfo
  {author} {\bibfnamefont {J.~K.}\ \bibnamefont {Wahlstrand}}, \ and\ \bibinfo
  {author} {\bibfnamefont {H.~M.}\ \bibnamefont {Milchberg}},\ }\href
  {http://dx.doi.org/10.1088/0953-4075/48/9/094011} {\bibfield  {journal}
  {\bibinfo  {journal} {Journal of Physics B}\ }\textbf {\bibinfo {volume}
  {48}},\ \bibinfo {pages} {094011} (\bibinfo {year} {2015})}\BibitemShut
  {NoStop}%
\bibitem [{\citenamefont {Th\'eberge}\ \emph {et~al.}(2006)\citenamefont
  {Th\'eberge}, \citenamefont {Liu}, \citenamefont {Simard}, \citenamefont
  {Becker},\ and\ \citenamefont {Chin}}]{Theberge2006}%
  \BibitemOpen
  \bibfield  {author} {\bibinfo {author} {\bibfnamefont {F.}~\bibnamefont
  {Th\'eberge}}, \bibinfo {author} {\bibfnamefont {W.}~\bibnamefont {Liu}},
  \bibinfo {author} {\bibfnamefont {P.~T.}\ \bibnamefont {Simard}}, \bibinfo
  {author} {\bibfnamefont {A.}~\bibnamefont {Becker}}, \ and\ \bibinfo {author}
  {\bibfnamefont {S.~L.}\ \bibnamefont {Chin}},\ }\href {\doibase
  10.1103/PhysRevE.74.036406} {\bibfield  {journal} {\bibinfo  {journal}
  {Physical Review E}\ }\textbf {\bibinfo {volume} {74}},\ \bibinfo {pages}
  {036406} (\bibinfo {year} {2006})}\BibitemShut {NoStop}%
\bibitem [{\citenamefont {Kiran}\ \emph
  {et~al.}(2010{\natexlab{b}})\citenamefont {Kiran}, \citenamefont {Bagchi},
  \citenamefont {Krishnan}, \citenamefont {Arnold}, \citenamefont {Kumar},\
  and\ \citenamefont {Couairon}}]{Kiran2010a}%
  \BibitemOpen
  \bibfield  {author} {\bibinfo {author} {\bibfnamefont {P.~P.}\ \bibnamefont
  {Kiran}}, \bibinfo {author} {\bibfnamefont {S.}~\bibnamefont {Bagchi}},
  \bibinfo {author} {\bibfnamefont {S.~R.}\ \bibnamefont {Krishnan}}, \bibinfo
  {author} {\bibfnamefont {C.~L.}\ \bibnamefont {Arnold}}, \bibinfo {author}
  {\bibfnamefont {G.~R.}\ \bibnamefont {Kumar}}, \ and\ \bibinfo {author}
  {\bibfnamefont {A.}~\bibnamefont {Couairon}},\ }\href
  {http://dx.doi.org/10.1103/PhysRevA.82.013805} {\bibfield  {journal}
  {\bibinfo  {journal} {Physical Review A}\ }\textbf {\bibinfo {volume} {82}},\
  \bibinfo {pages} {013805} (\bibinfo {year} {2010}{\natexlab{b}})}\BibitemShut
  {NoStop}%
\bibitem [{\citenamefont {M\'echain}\ \emph {et~al.}(2004)\citenamefont
  {M\'echain}, \citenamefont {Couairon}, \citenamefont {Franco}, \citenamefont
  {Prade},\ and\ \citenamefont {Mysyrowicz}}]{Mechain2004}%
  \BibitemOpen
  \bibfield  {author} {\bibinfo {author} {\bibfnamefont {G.}~\bibnamefont
  {M\'echain}}, \bibinfo {author} {\bibfnamefont {A.}~\bibnamefont {Couairon}},
  \bibinfo {author} {\bibfnamefont {M.}~\bibnamefont {Franco}}, \bibinfo
  {author} {\bibfnamefont {B.}~\bibnamefont {Prade}}, \ and\ \bibinfo {author}
  {\bibfnamefont {A.}~\bibnamefont {Mysyrowicz}},\ }\href {\doibase
  10.1103/PhysRevLett.93.035003} {\bibfield  {journal} {\bibinfo  {journal}
  {Physical Review Letters}\ }\textbf {\bibinfo {volume} {93}},\ \bibinfo
  {pages} {035003} (\bibinfo {year} {2004})}\BibitemShut {NoStop}%
\end{thebibliography}%

\end{document}